\documentclass{ar-1col}
\usepackage[numbers]{natbib}
\usepackage{bm}
\usepackage{graphicx}
\usepackage{mathrsfs}
\usepackage{amsmath}
\usepackage{amssymb}
\usepackage{relsize}

\newcommand{\rhozero}{\rho_{\raisebox{-2.0pt}{\tiny\!0}}}
\newcommand{\epszero}{\varepsilon_{\raisebox{-2.0pt}{\tiny\!0}}}

\def\yR{y_{\raisebox{-0.75pt}{\tiny {\rm R}}}}
\newcommand{\Edens}{\mathlarger{\mathlarger{\varepsilon}}}

\jname{Xxxx. Xxx. Xxx. Xxx.}
\jvol{AA}
\jyear{YYYY}
\doi{10.1146/((please add article doi))}

\begin{document}
\markboth{Yang and Piekarewicz}{Covariant Nuclear Density Functional Theory}

\title{Covariant Density Functional Theory in Nuclear Physics and Astrophysics}

\author{Junjie Yang and J. Piekarewicz
\affil{Department of Physics, Florida State University, Tallahassee, FL 32306-4350, U.S.A; email: jpiekarewicz@fsu.edu}}

\begin{abstract}   
\emph{How does subatomic matter organize itself?} Neutron stars are cosmic laboratories uniquely poised to answer this 
fundamental question that lies at the heart of nuclear science. Newly commissioned rare isotope facilities, telescopes 
operating across the entire electromagnetic spectrum, and ever more sensitive gravitational wave detectors will probe the
properties of neutron-rich matter with unprecedented precision over an enormous range of densities. Yet, a coordinated 
effort between observation, experiment, and theoretical research is of paramount importance for realizing the full potential 
of these investments. Theoretical nuclear physics provides valuable insights into the properties of neutron-rich matter in 
regimes that are not presently accessible to experiment or observation. In particular, nuclear density functional theory is 
likely the only tractable framework that can bridge the entire nuclear landscape by connecting finite nuclei to neutron 
stars. This compelling connection is the main scope of the present review.
\end{abstract}

\maketitle

\tableofcontents

\section{INTRODUCTION}

Nuclear science is poised to enter a period of transformational changes driven by the upgrade and commissioning
of state-of-the-art experimental and observational facilities. As we embark on this new journey of discovery, nuclear 
theory will play a critical role in guiding new experimental programs and in predicting the properties of nuclear matter 
in regimes 
that will remain inaccessible to experiment and observation. With unparalleled depth and breadth, nuclear science 
is driven by the quest to answer fundamental questions ranging from the quark-gluon structure of hadronic matter 
to the synthesis of heavy elements in cataclysmic stellar explosions\,\cite{LongRangePlan}. In this contribution we 
focus on the critical role that Density Functional Theory (DFT) plays in on our understanding of a variety of 
nuclear phenomena that range from the structure and dynamics of exotic nuclei to the fascinating properties of 
neutron stars. Remarkable advances in theoretical nuclear physics have propelled traditional wave function methods 
to such heights that highly accurate predictions of the properties of small to medium size nuclei are now routine; see
Refs.\,\cite{Pieper:2001mp,Barrett:2013nh,Hagen:2013nca,Navratil:2016ycn} and references contained therein. 
Such ``ab initio" approaches provide meaningful benchmarks for the development of reliable energy density functionals
which can then be applied to larger nuclear systems. Indeed, this powerful connection between ab initio approaches 
and DFT is one of the main motivations behind the ${}^{48}$Ca Radius EXperiment (CREX) at Jefferson 
Lab\,\cite{CREX:2013,Horowitz:2013wha}. Multiple paths exist for improving the performance of nuclear energy 
density functionals and on transforming them into proper effective field theories. For a recent perspective on how
to approach this challenging task see Ref.\,\cite{Furnstahl:2019lue} and references contained therein.

Density Functional Theory  is a powerful technique developed by Kohn and collaborators\,\cite{Hohenberg:1964zz,
Kohn:1965}  in the mid 60s to understand the electronic structure of complex many-body systems and for which Kohn 
was recognized with the 1998 Nobel Prize in Chemistry\,\cite{Kohn:1999}. Today, DFT is widely used in chemistry as
well as in many areas of physics\,\cite{Fiolhais:2003,Burke:2007,Skylaris,Drut:2009ce}.
In its original application to electronic structure, Hohenberg and Kohn (HK) assumed the 
validity of the Born-Oppenheimer approximation, which defines the many-body Hamiltonian in terms of a conventional 
kinetic-energy contribution, a two-body potential that accounts for the electronic repulsion, and a one-body attractive 
potential provided by the ``stationary" nuclei. Given that in the Born-Oppenheimer approximation the position of the 
heavy nucleus is assumed to be fixed, this last term is commonly referred to as the \emph{external} potential. DFT is 
firmly rooted in the two Hohenberg-Kohn theorems which state: (a) that a one-to-one correspondence exists 
between the one-body electronic density and a suitable external potential and (b) that an energy density functional 
(EDF) exists which upon functional minimization yields both the exact ground-state energy and one-body density of 
the complicated many-body system\,\cite{Hohenberg:1964zz}. Essentially, the HK theorems establish a remarkable 
and subtle result, namely, that the exact ground-state energy of the complicated many-body system may be obtained 
from minimizing a \emph{suitable} EDF that only depends on the one-body density. Perhaps the greatest virtue of DFT 
is that it shifts the focus from the complicated many-body wave function that depends on $3N$ spatial coordinates 
(for an $N$-particle system) to the much more intuitive one-body density that depends only on three. By doing so, 
DFT not only reduces drastically the complexity of the problem, but also invites physical insights into the construction 
of the functional. This is particularly relevant given that the HK theorem is an \emph{existence} theorem that 
offers no guidance on how to construct the appropriate energy density functional. This presents a serious challenge 
to the implementation, as no accurate representation of the kinetic energy part of the EDF exists.

In an effort to mitigate this problem and inspired by Hartree-Fock theory, Kohn and Sham replaced the complex 
interacting system by an equivalent system of \emph{non-interacting} electrons moving in a suitably-generated 
external potential\,\cite{Kohn:1965}. The term ``equivalent" is used to indicate that the Kohn-Sham (KS) potential 
must be sophisticated enough to reproduce the exact one-body density of the interacting system. So while the KS 
equations for the fictitious system closely resemble the structure of the Hartree equations, they differ 
by the presence of an \emph{exchange correlation} term that ensures that its density is identical to that of the 
interacting system. In essence, the KS approach trades the search for an accurate energy density functional for 
that of a complex exchange correlation potential. Nevertheless, the reformulation of the DFT problem in terms of 
one-particle orbitals has several advantages. First, unlike ``orbital-free" DFT where the kinetic-energy functional is 
unknown and complex, the kinetic energy term for the fictitious system  is known. Second, the computational cost 
is minimal as it increases linearly with the number of occupied orbitals. Third, the construction of the one-body 
density involves a simple sum over the occupied single-particle orbitals. Finally, self-consistent problems of this 
kind have been around for almost a century, so efficient and robust methods for their solution abound. Note that 
self-consistency is demanded because the one-body density depends on the single-particle orbitals which, in turn, 
are solutions of a Schr\"odinger (or Dirac) equation in the presence of a density-dependent KS potential. 

After this historical interlude it is appropriate to ask how can DFT be extended from the electronic sector to the nuclear 
domain. Unfortunately, the answer is far from obvious\,\cite{Furnstahl:2019lue}. One immediate difficulty concerns the 
one-to-one correspondence between the one-body density and the external potential, a concept that lies at the heart of 
DFT.  As a self-bound, many-body system, atomic nuclei are not subjected to any external potential. Hence, within the 
scope of the original ``orbital-free" DFT of Hohenberg and Kohn\,\cite{Hohenberg:1964zz}, the generalization to nuclear 
physics is unclear. Yet, within the ``mean-field like" Kohn-Sham paradigm some similarities emerge.  After all, mean-field 
theory has been an integral part of nuclear theory for many decades; see Ref.\,\cite{Negele:1981tw} and references 
contained therein. Although the external potential is germane to the KS formalism, could one simply  regard the nuclear 
mean field as the KS potential without the all important external potential? Regretfully, this is not the case, mainly 
because of the necessity of the complicated \emph{exchange-correlation} potential that is essential to reproduce the 
exact ground state energy. Indeed, neglecting the exchange correlation potential reduces the KS equations to the much 
simpler set of Hartree equations\,\cite{Kohn:1999}. However, in the context of nuclear physics it is well known that a 
Hartree potential computed from the convolution of the ``bare" nucleon-nucleon interaction with the nuclear density 
provides a poor description of the properties of atomic nuclei\,\cite{Negele:1981tw}.  To overcome this problem 
``effective density dependent forces" were developed by Skyrme almost a decade before the inception of density 
functional theory\,\cite{Skyrme:1956zz,Skyrme:1959zz}. In particular, part of the success of the Skyrme interaction 
relies on the existence of powerful relations connecting the (isoscalar) parameters of the model to various bulk properties 
of infinite nuclear matter, such as the saturation density, binding energy per nucleon, and incompressibility 
coefficient\,\cite{Vautherin:1971aw,Stone:2006fn}. In this manner important features of the nuclear dynamics are 
directly encoded into the parameters of the model. Reminiscent of the Hartree-Fock---or the more modern Kohn-Sham 
approach---the resulting single-particle equations of motion are derived from functional minimization of a properly 
defined Skyrme energy density functional. So while the notion of a nuclear mean-field potential remains essential, 
its connection to the underlying (or ``bare") nucleon-nucleon interaction has been lost. Indeed, present day nuclear 
EDFs are largely empirical, as the parameters of the model have no direct connection to the underlying nucleon-nucleon
interaction often calibrated using deuteron properties and phase shifts. Rather, in DFT the model parameters are fitted 
to selected properties of atomic nuclei. One often justifies empirical EDFs by invoking the HK theorems, which as 
existence theorems provide no guidance on how to construct the functional. Nevertheless, significant advances have 
been made over the last decade to mitigate the reliance on empirical EDFs in favor of more fundamental ones; for an
extensive reviews entitled \emph{Toward ab initio density functional theory for nuclei} see Ref.\,\cite{Drut:2009ce} and 
references contain therein. In parallel, much effort has also been devoted to the construction of a \emph{Universal Nuclear 
Energy Density Functional} with the aim of achieving a comprehensive understanding of finite 
nuclei across the entire nuclear landscape\,\cite{Bertsch:2007,Stoitsov:2007,Kortelainen:2010hv,Erler:2012,Kortelainen:2013}. 

In this review we will continue to rely on empirical EDFs, but within the context of \emph{covariant density functional theory}. 
Our motivation for this generalization is mostly pragmatical, as we seek a unified approach that can simultaneously describe 
the dynamics of finite nuclei and neutron stars, systems with natural length scales that differ by 18 orders of magnitude! We 
aim to build high-quality functionals that yield an accurate description of the properties of finite nuclei, generate an equation 
of state that is consistent with known neutron-star properties, while providing a Lorentz covariant extrapolation to dense matter. 
In the case of finite nuclei, an important goal is not only to compute ground state properties, but also the linear response of the 
ground state to a variety of probes. In this context DFT continues to provide an ideal framework. Indeed, given the variational 
nature of DFT, small oscillations around the variational minimum encapsulate the linear response of the ground state to weak 
external perturbations. However, care must be exercised in employing a residual interaction that is consistent with the one 
employed to generate the ground state. Only then can one ensure that important symmetries and conservation laws are properly
enforced\,\cite{Thouless:1961,Dawson:1990wp,Piekarewicz:2001nm,Ring:2004}. Finally, given that some of the observables 
of interest require extrapolations into regions that are inaccessible in the laboratory, we aim when possible to supplement  our
predictions with theoretical uncertainties\,\cite{Kortelainen:2010hv,Fattoyev:2011ns,Reinhard:2010wz,Fattoyev:2011ns,
Fattoyev:2012rm,Reinhard:2012vw,Reinhard:2013fpa,Dobaczewski:2014jga,Piekarewicz:2014kza,Chen:2014sca}. This can
now be done routinely as the calibration of the EDF produces a statistically robust covariant matrix.

Exploring the synergy between nuclear physics and astrophysics has always been a core mission of nuclear science.
In the particular context of neutron stars, the equation of state prescribed by the underlying DFT becomes essential in 
the description of the structure and dynamics of these fascinating compact objects. The powerful connection between 
nuclear physics and astrophysics has just been strengthen even further with the first direct detection of gravitational waves 
from the binary neutron star merger GW170817\,\cite{Abbott:PRL2017}. In one clean sweep GW170817 has 
confirmed the long-held belief that short gamma ray burst are associated with the merger of  two neutron stars; has 
identified the left-over \emph{kilonova} as the electromagnetic transient powered by the radioactive decay of the heavy 
elements synthesized in the rapid neutron-capture process\,\cite{Drout:2017ijr,Cowperthwaite:2017dyu,Chornock:2017sdf,
Nicholl:2017ahq}; and has provided stringent constraints on the equation of state\,\cite{Bauswein:2017vtn,
Fattoyev:2017jql,Annala:2017llu,Abbott:2018exr,Most:2018hfd,Tews:2018chv,Malik:2018zcf,Tsang:2018kqj,Radice:2018ozg}. 
Assessing the impact of this historic discovery will be an important component of this review.

We have organized the review as follows. In Sec.\,\ref{sec:Formalism} we start by discussing the class of covariant 
density functionals that will be considered in this work. We then introduce the associated set of  equations that must
be solved to obtain Kohn-Sham orbitals and ground-state densities. We then proceed to illustrate, also in 
Sec.\,\ref{sec:Formalism}, how to compute the nuclear matter equation of state using the same exact covariant EDFs. 
Note that the EOS is the sole ingredient required to solve the equations of hydrostatic equilibrium from which several 
neutron-star properties are extracted. Having developed the formalism, we then move to Sec.\,\ref{sec:Results} where 
our predictions are discussed, with special emphasis on those observables that are difficult to probe under present 
laboratory conditions, either because of the large neutron excess or the very high density. We conclude and offer our
perspectives for the future in Sec.\,\ref{sec:Conclusions}.

\section{FORMALISM}
\label{sec:Formalism}

In this section we develop the formalism underpinning covariant density functional theory and focus on its
implementation to the physics of finite nuclei and neutron stars. The Dirac equation obeyed by the nucleon  
fields and the associated Klein-Gordon equations for the meson fields may be regarded as the generalization 
of the Kohn-Sham equations to the domain of covariant DFT. Note that as alluded earlier, the effective interaction 
bares little resemblance to the underlying nucleon-nucleon interaction, as the parameters of the model
are calibrated to the properties of finite nuclei rather than to two-nucleon data. The application to neutron stars
relies on the same energy density functional without any adjustments. That is, the equation of state that serves 
as the sole input for the Tolman-Oppenheimer-Volkoff equations is constructed from the same model used to
compute the properties of finite nuclei, thereby connecting problems with length scales that differ by about 18
orders of magnitude. Because of space limitations, we omit discussing the collective nuclear response, an 
interesting area of investigation that will continue to thrive with the advent of radioactive beam facilities. For
a review of collective excitations in the context of covariant DFT see 
Refs.\,\cite{Piekarewicz:2013bea,Piekarewicz:2015yla} and references contained therein.

\subsection{Covariant Density Functional Theory}
\label{sec:DFT} 

Finite nuclei are complex many-body systems governed largely by the strong nuclear force. Although quantum 
chromodynamics (QCD) is the fundamental theory of the strong interaction, many technical hurdles still prevent 
us from applying QCD in the non-perturbative regime of relevance to nuclear physics. To date, density functional
theory is the most promising---and perhaps only tractable---approach that may be applied over the entire nuclear 
landscape: from finite nuclei to neutron stars. In the traditional non-relativistic approach, the dynamical information 
is encoded in an effective interaction between nucleons that is used to build the energy density functional in terms 
of conserved isoscalar and isovector (or proton and neutron) densities and their associated 
currents\,\cite{Bertsch:2007,Stoitsov:2007}. The paradigm of such an effective non-relativistic interaction is the 
Skyrme interaction\,\cite{Skyrme:1956zz,Skyrme:1959zz,Vautherin:1971aw,Stone:2006fn}. Given that the model 
parameters cannot be computed from first principles, various optimization protocols are being used to adjust 
their values by fitting to a suitable set of experimental data\,\cite{Kortelainen:2010hv,Erler:2012,Kortelainen:2013}. 
From such an optimally calibrated density functional, one derives the corresponding Kohn-Sham equations which 
are then solved using self-consistent mean-field methods\,\cite{Bender:2003jk}

Covariant density functional theory follows in the footsteps of Skyrme DFT, but with both nucleons and mesons 
as the fundamental degrees of freedom. Among the earliest attempts at a relativistic description of the nuclear 
dynamics is the work of Johnson and Teller\,\cite{Johnson:1955zz}, Duerr\,,\cite{Duerr:1956zz}, and Miller and 
Green\,\cite{Miller:1972zza}; for a more complete historical account see Ref.\,\cite{Serot:1984ey}. Besides a
desire to understand the saturation of nuclear matter and its impact on the ground-state energy and densities 
of atomic nuclei, an important motivation for a relativistic description---and one that remains true to this day---was 
the development of a theory of highly condensed matter that could be applied to the study of neutron 
stars\,\cite{Walecka:1974qa}. Originally, Quantum HadroDynamics (or QHD) was conceived as a quantum field 
theory consisting of a nucleon field interacting via the exchange of neutral scalar and vector 
mesons\,\cite{Walecka:1974qa}. Remarkably, a self-consistently generated equation of state for symmetric 
nuclear matter exhibits saturation---even at the mean field level---because of the different Lorentz character
of the scalar and vector interactions. Moreover, pure neutron matter was found to be unbound and to remain 
causal at all densities. However, whereas nuclear saturation---the existence of an equilibrium density at which 
the pressure vanishes---represented a great triumph of the theory, the curvature around the minimum, {\sl i.e.,}
the incompressibility coefficient, was inconsistent with experimental limits obtained from the measurement of the 
monopole response of heavy nuclei\,\cite{Youngblood:1977}. To remedy this deficiency, scalar-meson self 
interactions, first introduced by Boguta and Bodmer\,\cite{Boguta:1977xi}, were successful in softening the 
equation of state. Since then, modifications to the underlying effective Lagrangian density were introduced 
in an effort to provide a more accurate description of the properties of finite nuclei and neutron 
stars\,\cite{Chen:2014sca,Serot:1984ey,Mueller:1996pm,Lalazissis:1996rd,Vretenar:2003qm,
Lalazissis:2005de,Todd-Rutel:2005fa,Agrawal:2010wg,Fattoyev:2010mx,Chen:2014mza}. Moreover, some 
of the most recent parametrizations now provide properly quantified statistical uncertainties.

In the framework of covariant DFT, the basic degrees of freedom are the nucleon (protons and neutrons), 
three mesons, and the photon. The isodoublet nucleon field $\psi$ interacts via the exchange of photons 
($A_\mu$)  as well as three massive ``mesons": the isoscalar-scalar $\sigma$ meson, the isoscalar-vector 
$\omega$ meson, and the isovector-vector $\rho$ meson \cite{Serot:1984ey,Walecka:1974qa, Serot:1997xg}. 
The effective (interacting) Lagrangian density takes the following form\,\cite{Serot:1984ey,
Mueller:1996pm,Horowitz:2000xj,Todd:2003xs},
\begin{eqnarray}
{\mathscr L}_{\rm int} &=&
\bar\psi \left[g_{\rm s}\phi   \!-\! 
         \left(g_{\rm v}V_\mu  \!+\!
    \frac{g_{\rho}}{2}{\mbox{\boldmath$\tau$}}\cdot{\bf b}_{\mu} 
                               \!+\!    
    \frac{e}{2}(1\!+\!\tau_{3})A_{\mu}\right)\gamma^{\mu}
         \right]\psi \nonumber \\
                   &-& 
    \frac{\kappa}{3!} (g_{\rm s}\phi)^3 \!-\!
    \frac{\lambda}{4!}(g_{\rm s}\phi)^4 \!+\!
    \frac{\zeta}{4!}   g_{\rm v}^4(V_{\mu}V^\mu)^2 +
   \Lambda_{\rm v}\Big(g_{\rho}^{2}\,{\bf b}_{\mu}\cdot{\bf b}^{\mu}\Big)
                           \Big(g_{\rm v}^{2}V_{\nu}V^{\nu}\Big) \nonumber \\ 
   &\equiv& \bar\psi \left[g_{\rm s}\phi   \!-\! 
         \left(g_{\rm v}V_\mu  \!+\!
    \frac{g_{\rho}}{2}{\mbox{\boldmath $\tau$}}\cdot{\bf b}_{\mu} 
                               \!+\!    
    \frac{e}{2}(1\!+\!\tau_{3})A_{\mu}\right)\gamma^{\mu}
         \right]\psi - U_{\rm eff}(\phi,V_{\mu},{\bf b}_{\mu}).                                                   
 \label{LDensity}
\end{eqnarray}
The first line in the above equation contains the conventional meson-nucleon Yukawa couplings, while
the second line includes non-linear meson interactions $U_{\rm eff}(\phi,V_{\mu},{\bf b}_{\mu})$ that 
serve to simulate the complicated many-body dynamics and that are required to improve the predictive 
power of the model. As already alluded, the two isoscalar parameters $\kappa$ and $\lambda$ introduced 
by Boguta and Bodmer\,\cite{Boguta:1977xi} were designed to reduce the incompressibility coefficient of 
symmetric nuclear matter in accordance to  measurements of giant monopole resonances in finite 
nuclei\,\cite{Youngblood:1977}. Sometime later, M\"uller and Serot introduced the isoscalar parameter 
$\zeta$ to soften the equation of symmetric nuclear matter but at much higher densities\,\cite{Mueller:1996pm}. 
Indeed, they found that by tuning the value of $\zeta$ one could significantly modify the maximum neutron star 
mass without compromising the success of the model in reproducing ground-state observables. Finally, 
the mixed isoscalar-isovector parameter $\Lambda_{\rm v}$ was introduced to modify the density dependence 
of the symmetry energy---particularly its slope at saturation density $L$. The structure of both neutron-rich 
nuclei and neutron stars is highly sensitive to the slope of the symmetry energy\,\cite{Horowitz:2000xj,
Horowitz:2001ya,Carriere:2002bx}. 

The field equations resulting from the above Lagrangian density may be solved exactly in the mean-field 
limit, where the meson-field operators are replaced by their classical expectation values\,\cite{Serot:1984ey,
Walecka:1974qa}. For a static and spherically symmetric ground state this implies:
\begin{subequations}
\begin{align}
\phi(x)  \rightarrow & \langle\phi(x)\rangle=\phi_{0}(r),\\
V^{\mu}(x)  \rightarrow & \langle V^{\mu}(x)\rangle=g^{\mu0}V_{0}(r),\\
b_{a}^{\mu}(x)  \rightarrow & \langle b_{a}^{\mu}(x)\rangle=g^{\mu_{0}}\delta_{a3}b_{0}(r),\\
A^{\mu}(x)  \rightarrow & \langle A^{\mu}(x)\rangle=g^{\mu0}A_{0}(r).
\end{align}
\end{subequations}

Given that the meson fields couple to their associated bilinear nucleon currents, the baryon sources must 
also be replaced by their (normal-ordered) expectation values in the mean-field ground state:
\begin{subequations}
\begin{align}
\bar{\psi}(x)\psi(x)  & \rightarrow  \langle:\!\bar{\psi}(x)\psi(x)\!:\rangle = \rho_{\rm s}(r),\\
\bar{\psi}(x)\gamma^{\mu}\psi(x) & \rightarrow \langle:\!\bar{\psi(x)}\gamma^{\mu}\psi(x)\!:\rangle = g^{\mu0}\rho_{\rm v}(r), \\
\bar{\psi}(x)\gamma^{\mu}\tau_{a}\psi(x) & \rightarrow \langle:\!\bar{\psi(x)}\gamma^{\mu}\tau_{a}\psi(x)\!:\rangle = g^{\mu0}\delta_{a3}\rho_{3}(r), \\
\bar{\psi}(x)\gamma^{\mu}\tau_{p}\psi(x) & \rightarrow \langle:\!\bar{\psi(x)}\gamma^{\mu}\tau_{p}\psi(x)\!:\rangle = g^{\mu0}\rho_{p}(r), 
\end{align}
\end{subequations}
where $\rho_{\rm s}$ is the dynamically generated scalar density, $\rho_{\rm v}$ the conserved isoscalar baryon density, 
$\rho_3$ the isovector baryon density, and $\rho_p$ the proton density. In terms of the individual proton and neutron 
densities, one can write $\rho_{\rm v}\!=\!\rho_{p}\!+\!\rho_{n}$ and $\rho_3\!=\!\rho_{p}\!-\!\rho_{n}$. Note that we have
introduced the proton isospin projection operator as $\tau_{p}\!=\!(1\!+\!\tau_{3})/2$. Using the above approximations one 
can now derive the associated Euler-Lagrangian equations of motion for a generic quantum field $q_{i}$\,\cite{Serot:1984ey}:
\begin{align}
\partial_{\mu}\left[\frac{\partial\mathscr{L}}{\partial\left(\partial_{\mu}q_{i}\right)}\right] - \frac{\partial \mathscr{L}}{\partial q_{i}} = 0.
\end{align}

In the particular case of the Lagrangian density given in Eq.(\ref{LDensity}), the classical meson field satisfy Klein-Gordon 
equations containing both non-linear meson interactions and ground-state baryon densities as source terms. That is,
\begin{subequations}
\begin{align}
\Big(\nabla^{2} - m_{s}^{2}\Big)\phi_{0}(r) - \frac{\partial U_{\text{eff}}}{\partial \phi_0} = & -g_{\rm s} \rho_{\rm s}(r),\\
\Big(\nabla^{2} - m_{v}^{2}\Big)V_{0}(r) + \frac{\partial U_{\text{eff}}}{\partial V_0}= &-g_{\rm v}\rho_{\rm v}(r),\\
\Big(\nabla^{2} - m_{\rho}^{2}\Big)b_{0}(r) + \frac{\partial U_{\text{eff}}}{\partial b_0} = & -\frac{g_{\rho}}{2} \rho_{3}(r).
\end{align}
\label{KGEqns}
\end{subequations} 
In turn, the Coulomb field satisfied the much simpler Poisson's equation:
\begin{equation}
\nabla^{2}A_{0} =  -e\rho_{p}.
\label{PoissonEqn}
\end{equation} 
On the other hand, the nucleons satisfy a Dirac equation with the meson fields generating scalar and time-like vector 
mean-field potentials. That is,
\begin{equation}
\left[-i{\bm\alpha}\cdot{\bm\nabla} + g_{\rm v}V_0(r) + \frac{g_{\rho}}{2} \tau_3 b_0(r) + e\tau_{p} A_0(r) 
 +\beta\big(M-g_{\rm s} \phi_0(r)\big)\right]\psi({\bf r})  = E\psi({\bf r}).
 \label{DiracEqn}
\end{equation}
The above set of equations---Eqs.(\ref{KGEqns}-\ref{DiracEqn})---represent the effective Kohn-Sham equations 
for the nuclear many-body problem. As such, this set of mean-field equations must be solved self-consistently.
That is, the single-particle orbitals satisfying the Dirac equation are generated from the various meson fields 
which, in turn, satisfy Klein-Gordon equations with the appropriate ground-state densities as the source terms. 
This demands an iterative procedure in which mean-field potentials of the Wood-Saxon form are initially provided 
to solve the Dirac equation for the occupied nucleon orbitals which are then combined to generate the appropriate 
densities for the meson field. The Klein-Gordon equations are then solved with the resulting meson fields providing
a refinement to the initial mean-field potentials. This procedure continues until self-consistency is achieved; see 
Ref.\cite{Todd:2003xs} for a detailed description on the implementation. Due to the highly non-linear structure of 
these equations, extra care must be exercised in ensuring that self-consistency has indeed been achieved. 

In the spirit of covariant DFT, the outcome of the iterative procedure are ground-state densities, binding energies,
and self-consistent mean fields. However, given the empirical nature of the covariant DFT, one must first adjust the 
parameters of the interacting density given in Eq.(\ref{LDensity}) to available experimental/observational data. 
Recently, such calibrating procedure has been implemented without any reliance on ``pseudo-data", namely without 
incorporating assumed bulk properties of infinite nuclear matter\,\cite{Chen:2014sca,Chen:2014mza}. Moreover, 
besides predicting (rather than assuming) the values of several bulk properties of nuclear matter, the statistical 
approach adopted in the calibrating procedure allows one to provide quantifiable theoretical errors. In doing so, 
one discovers that the isoscalar sector of the density functional, namely, the sector that does not distinguish 
neutrons from protons, is fairly well constrained by existing nuclear observables. This is not surprising as most of 
the experimental nuclear observables available today probe small to moderate neutron-proton asymmetries. In 
contrast, the isovector sector of the nuclear density functional is poorly constrained. As it stands now, the two 
isovector parameters defining the effective Lagrangian density in Eq.(\ref{LDensity}) are the Yukawa coupling 
$g_{\rho}$ and the mixed isoscalar-isovector coupling $\Lambda_{\rm v}$. As shown in Ref.\,\cite{Chen:2014sca}, 
these two model parameters can be fixed once two fundamental parameters of the nuclear symmetry energy are 
inferred; see Sec.\ref{sec:EOS}. Enormous theoretical and experimental efforts have been devoted for the last 
two decades to constrain these two parameters, or more generally the density dependence of the nuclear symmetry 
energy. Progress towards achieving this goal by using both laboratory data and astrophysical observables is an 
important component of this review.  
 
\subsection{Neutron Stars}
\label{sec:NS}

Having explained the main features of the covariant DFT formalism, we are now in a position to examine the
structure of neutron stars. The structure of spherically symmetric neutron stars in hydrostatic equilibrium---in 
particular the fundamental mass-vs-radius relation---is encapsulated in the Tolman-Oppenheimer-Volkoff (TOV) 
equations\,\cite{Tol39_PR55,Opp39_PR55}. Adopting natural units in which $G\!=\!c\!=\!1$, the TOV equation 
are given by
\begin{subequations}
 \begin{align}
  & \frac{dP(r)}{dr} = -\frac{\Big(\Edens(r)+P(r)\Big)
      \left(M(r)+4\pi r^{3}P(r)\right)}
      {r^{2}\Big(1-2M(r)/r\Big)}\,, \\
  & \frac{dM(r)}{dr} = 4\pi r^{2}\Edens(r)\,.
  \end{align}
 \label{TOV}
\end{subequations}
Here $M(r)$, $P(r)$, $\Edens(r)$ represent the enclosed mass, pressure, and energy density profiles, 
respectively.  The TOV equations represent the extension of Newtonian gravity to the domain of general 
relativity. Such extension is essential as the typical escape velocity from the surface of neutron star is 
close to the speed of light. Indeed, the Schwarzschild radius of a neutron star (of the order of 3-6 kilometers) 
is comparable to its 12-14 kilometer radius. That is,
\begin{equation}
 \frac{v_{\rm esc}^{2}}{c^{2}} = \frac{2GM}{c^{2}R} \equiv \frac{R_{\rm s}(M)}{R}.
 \label{Vescape}
\end{equation}
Upon inspection, one notices that the only input required for the solution of the TOV equations is an equation 
of state, namely, a relation between the pressure and the energy density. Providing such an EOS is within the 
purview of nuclear physics. Although unknown to Oppenheimer and Volkoff at the time of their original
contribution\,\cite{Opp39_PR55}, the main reason that nuclear physics plays such a predominant role is 
easy to understand. Back in 1939 Oppenheimer and Volkoff concluded that a neutron star supported exclusively 
by the quantum pressure from its degenerate neutrons will collapse once its mass exceeds $0.7\,M_{\odot}$. 
Today, however, the evidence for significantly more massive neutron stars is 
overwhelming\,\cite{Thorsett:1998uc,Lattimer:2004pg}. Indeed, within the last 
decade the existence of three neutron stars with masses in the vicinity of 2$M_{\odot}$ has been firmly 
established\,\cite{Demorest:2010bx,Antoniadis:2013pzd,Cromartie:2019kug}. In fact, the most massive neutron 
star observed to date ($M\!=\!2.1^{+0.10}_{-0.09}\,M_{\odot}$) was reported very recently by Cromartie and 
collaborators\,\cite{Cromartie:2019kug}. This implies that the additional support against gravitational collapse 
must come from nuclear interactions, which at the high densities (or short distances) of the stellar core are known 
to be strongly repulsive. The large discrepancy between recent observations and the 80-year old prediction by 
Oppenheimer and Volkoff has effectively transferred ownership of the neutron-star problem to nuclear physics. It 
is appropriate to mention that unlike the well-known collapse of a white-dwarf star, the existence of a maximum 
neutron-star mass is a purely general-relativistic effect with no counterpart in Newtonian gravity. Whereas the 
collapse of a white-dwarf star is characterized by a dramatic reduction in the stellar radius as the mass approaches 
the Chandrasekhar limit of $M_{\rm Ch}\!=\!1.4\,M_{\odot}$\,\cite{Chandrasekhar:1931}, the existence of a maximum 
neutron-star mass develops as an instability against small radial perturbations\,\cite{Weinberg:1972}. The maximum 
neutron-star mass is presently unknown, although it has been suggested that GW170817 already provides some 
important constraints\,\cite{Margalit:2017dij}. 

The existence of neutron stars with masses in excess of $2M_{\odot}$ demands a ``stiff'' equation of state, 
namely, one in which the pressure increases rapidly with energy density.
In contrast, the recent detection of gravitational waves from the binary neutron star merger GW170817 suggests 
that the equation of state must be soft. This conclusion was drawn based on the extraction of a rather small value 
for the tidal deformability (or polarizability) of a $M\!=\!1.4\,M_{\odot}$ neutron star\,\cite{Abbott:PRL2017,Abbott:2018exr}. 
The dimensionless tidal deformability is defined as
\begin{equation}
 \Lambda = \frac{2}{3}k_{2}\left(\frac{c^{2}R}{GM}\right)^{\!5}
                 =\frac{64}{3}k_{2}\left(\frac{R}{R_{s}}\right)^{\!5},
 \label{Lambda}
\end{equation}
where $k_{2}$ is known as the second Love number\,\cite{Binnington:2009bb,Damour:2012yf}. Clearly, $\Lambda$ 
is extremely sensitive to the compactness parameter $\xi\!\equiv\!R_{s}/R$\,\cite{Hinderer:2007mb,Hinderer:2009ca,
Damour:2009vw,Postnikov:2010yn,Fattoyev:2012uu,Steiner:2014pda,Piekarewicz:2018sgy}. Given that $k_{2}$ is
known to display a mild sensitivity to the underlying equation of state\,\cite{Piekarewicz:2018sgy}, a measurement of 
$\Lambda$, for a given mass, determines the stellar radius and ultimately the stiffness of the equation of state.

Trying to account for both large masses and small radii creates an interesting tension that once resolved is bound 
to provide fundamental insights into the EOS. One possibility is that the equation of state is relatively soft at about 
twice nuclear matter saturation density, which is the region believed to be most strongly correlated to the stellar 
radius\,\cite{Lattimer:2006xb}. In this density domain the stellar radius is primarily controlled by the density dependence 
of the nuclear symmetry energy\,\cite{Fattoyev:2012rm,Chen:2014sca,Horowitz:2000xj,Horowitz:2001ya,
Carriere:2002bx}. On the other hand, the maximum stellar mass is controlled by the equation of state at the highest
densities. Thus, one may be able to account for both large masses and small radii if the equation of state is soft at
intermediate densities and then stiffens at higher densities. Insights into the behavior of the symmetry energy can
be gleaned from the recently completed (and currently being analyzed) PREX-II measurement of the neutron skin 
thickness of ${}^{208}$Pb at the Jefferson Laboratory. It has been demonstrated that the neutron skin thickness of 
${}^{208}$Pb is strongly correlated to the slope of the nuclear symmetry energy at saturation 
density\,\cite{Brown:2000,Furnstahl:2001un,Centelles:2008vu,RocaMaza:2011pm}.

Having established the importance of the tidal polarizability in elucidating the structure of neutron stars, we conclude 
this section with a brief description of the necessary steps involved in its computation. For simplicity, one can assume
that mass, pressure, and energy density profiles are available after having solved the TOV equations, leaving the 
second Love number $k_{2}$\,\cite{Binnington:2009bb,Damour:2012yf} as the only unknown parameter appearing in 
Eq.\,(\ref{Lambda}). Evidently, $\Lambda$ is extremely sensitive to the compactness parameter 
$\xi$\,\cite{Hinderer:2007mb,Hinderer:2009ca,Damour:2009vw,Postnikov:2010yn,
Fattoyev:2012uu,Steiner:2014pda}. In turn, the second Love number $k_{2}$ depends on both $\xi$ and 
$\yR$,
\begin{align}
 k_{2}(\xi,\yR) &= \frac{1}{20}\xi^{5}(1-\xi)^{2}\Big[(2\!-\!\yR)+(\yR\!-\!1)\xi\Big] \nonumber \\
                       &  \times \Bigg\{ \Big[(6\!-\!3\yR)+\frac{3}{2}(5\yR\!-\!8)\xi\Big]\xi 
                        + \frac{1}{2}\Big[(13\!-\!11\yR)+\frac{1}{2}(3\yR\!-\!2)\xi+\frac{1}{2}(1\!+\!\yR)\xi^{2}\Big]\xi^{3}\nonumber \\
                       &\hspace{10pt}+3\Big[(2\!-\!\yR)+(\yR\!-\!1)\xi\Big](1-\xi)^{2}\ln(1-\xi)\Bigg\}^{-1},
 \label{k2}
\end{align}
where $\yR\!\equiv\!y(r\!=\!R)$ is obtained after solving the following non-linear, first order differential equation for 
$y(r)$---a quantity that is closely related to the tidally-induced quadrupole 
field\,\cite{Postnikov:2010yn,Fattoyev:2012uu,Piekarewicz:2018sgy}. That is,
\begin{equation}
 r\frac{dy(r)}{dr} + y^{2}(r) + F(r)y(r) +r^{2}Q(r) = 0, \hspace{10pt} {\rm with}\; y(0)=2.
 \label{yofr}
\end{equation}
Here the two functions $F(r)$ and $Q(r)$ depend on the known mass, pressure, and energy density profiles of the 
star:
\begin{subequations}
 \begin{align}
  & F(r)  \!=\! \frac{1-4\pi r^{2}\Big(\Edens(r)-P(r)\Big)}
                {\displaystyle{\left(1\!-\!\frac{2M(r)}{r}\right)}},  \\  
  & Q(r) \!=\!  \frac{4\pi}{\displaystyle{\left(1\!-\!\frac{2M(r)}{r}\right)}} 
                \!\left(5\Edens(r)\!+\!9P(r)\!+\!\displaystyle{\frac{\Edens(r)\!+\!P(r)}{\mathlarger{c}_{\rm s}^{2}(r)}}
                \!-\!\frac{6}{4\pi r^{2}}\right) 
                \!-\!4\left[\frac{\Big(M(r)+4\pi r^{3}P(r)\Big)}{\displaystyle{r^{2}\left(1\!-\!\frac{2M(r)}{r}\right)}}\right]^{2}.
  \label{FandQ}
\end{align}  
\end{subequations}
Note that in addition $Q(r)$ depends on the speed of sound profile, which involves the derivative of the pressure with 
respect to the energy density, {\sl i.e.,}
\begin{equation}
 c_{\rm s}^{2}(r)\!=\!\frac{dP(r)}{d\Edens(r)}.
 \label{cs2}
 \end{equation}
A covariant energy density functional---unlike nonrelativistic functionals---ensures that the EOS remains causal
at all densities, namely, that the speed of sound never exceeds the speed of light.

\subsection{Equation of State}
\label{sec:EOS}
 
Neutron stars are ``cold" dense objects with a characteristic core temperature significantly lower than the 
corresponding Fermi temperature\,\cite{Lattimer:2006xb,Page:2006ud}. As such, and under the assumption
of spherical symmetry and hydrostatic equilibrium, the relevant equation of state is that of a zero temperature, 
electrically-neutral system in chemical (or ``beta") equilibrium. As
we aim to build a covariant energy density functional that describes the properties of both finite nuclei and
neutron stars, we adopt as the basic constituents of matter, neutrons, protons, and leptons (both electrons 
and muons). Note that leptons help maintain both charge neutrality and beta equilibrium, which ultimately
sets the proton fraction in the neutron star, a critical property that impacts many stellar properties. 

Although beta equilibrium dictates that only the total baryon density is conserved, we start with a discussion 
of the EOS of infinite nuclear matter where both neutron and proton densities are individually conserved. 
Infinite nuclear matter is an idealized system of protons and neutrons interacting solely via the strong nuclear 
force, so that both electromagnetic and weak interactions are ``turned off''. In such an idealized situation and
under the assumption of translational invariance, the expectation value of the various meson fields in 
Eq.(\ref{KGEqns}) are uniform (i.e., constant throughout space) and the Kohn-Sham orbitals in Eq.(\ref{DiracEqn}) 
are plane-wave Dirac spinors with medium-modified effective masses and energies that must be determined 
self-consistently. To derive the equation of state of infinite nuclear matter one invokes the energy-momentum 
tensor:
\begin{equation}
 {\mathscr T}_{\mu \nu} = - g_{\mu \nu}{\mathscr L} + 
 \left[\frac{\partial\mathscr{L}}{\partial\left(\partial^{\mu}q_{i}\right)}\right]\!\partial_\nu q_{i},
\label{Tmunu0}
\end{equation}
where a sum over all constituent fields $q_{i}$ is assumed. For a uniform system such as infinite nuclear
matter, the expectation value of the energy momentum tensor takes the following simple form\,\cite{Serot:1984ey}:
\begin{equation}
 \langle{\mathscr T}_{\mu \nu}\rangle = \big( \Edens + P\big) u_{\mu}u_{\nu} - P\,g_{\mu \nu},
\label{Tmunu1}
\end{equation}
where $u^{\mu}\!=\!\gamma(1,{\bm\beta})$ is the scaled four-velocity of the fluid that satisfies the 
Lorentz-invariant condition $u^{2}\!=\!u^{\mu}u_{\mu}=1$, with $\gamma$ being the Lorentz factor. 
In particular, for infinite nuclear matter at rest, {\sl i.e.,} $u^{\mu}\!=\!(1,{\bf 0})$, it follows that
\begin{equation}
  \Edens = \langle{\mathscr T}_{00}\rangle \hspace{5pt}{\rm and}\hspace{5pt}
  P = \frac{1}{3} \langle{\mathscr T}_{ii}\rangle.
\label{Tmunu2}
\end{equation}
Given that both the proton and neutron densities are conserved in infinite nuclear matter, 
the equation of state at zero temperature may be written as either a function of the
individual densities or as a function of the total baryon density $\rho\!=\rho_{n}\!+\!\rho_{p}$ 
and the neutron-proton asymmetry
$\alpha\!\equiv\!(\rho_{n}\!-\!\rho_{p})/(\rho_{n}\!+\!\rho_{p})$. Expanding the energy per
nucleon in even powers of the neutron-proton asymmetry is particularly insightful. That is,
\begin{equation}
  \frac{E}{A}(\rho,\alpha) -\!M \equiv {\cal E}(\rho,\alpha)
                          = {\cal E}_{\rm SNM}(\rho)
                          + \alpha^{2}{\cal S}(\rho)  
                          + {\cal O}(\alpha^{4}) \,.
 \label{EOS}
\end {equation}
Here ${\cal E}_{\rm SNM}(\rho)\!=\!{\cal E}(\rho,\alpha\!\equiv\!0)$ is the energy per nucleon 
of symmetric nuclear matter (SNM) and the symmetry energy ${\cal S}(\rho)$ represents the 
first-order correction to the symmetric limit. Note that no odd powers of $\alpha$ appear in
the expansion since in the absence of electroweak interactions the nuclear force is assumed 
to be isospin symmetric; isospin violations in the nucleon-nucleon interactions (which are small)
are henceforth neglected. Although there is \emph{a priori} no reason to neglect the higher order 
terms in Eq.(\ref{EOS}), for the models considered in this review the symmetry energy represents 
to a very good approximation the energy cost required to convert symmetric nuclear matter into 
pure neutron matter (PNM). That is,  
\begin{equation}
 {\cal S}(\rho)\!\approx\!{\cal E}(\rho,\alpha\!=\!1) \!-\! 
 {\cal E}(\rho,\alpha\!=\!0) \;.
 \label{SymmE}
\end {equation}
While the above relation is model dependent, its validity is easily verified in the case that protons 
and neutrons behave as non-interacting Fermi gases\,\cite{Piekarewicz2017}. 
The separation of the energy per nucleon as in Eq.(\ref{EOS}) is useful because symmetric nuclear 
matter is sensitive to the isoscalar sector of the density functional which is well constrained by the 
properties of stable nuclei. In contrast, the symmetry energy probes the isovector sector of the density 
functional which at present is poorly constrained because of the lack of experimental data on very 
neutron-rich systems. However, this problem will soon be mitigated with the commissioning of 
radioactive beam facilities throughout the world. 

Besides the separation of the EOS into isoscalar and isovector components, it is also useful to  
characterize the equation of state in terms of a few of its bulk parameters defined at saturation
density. Nuclear saturation, the existence of an equilibrium density that characterizes the interior 
of medium to heavy nuclei, is a hallmark of the nuclear dynamics. By performing a Taylor series 
expansion around nuclear matter saturation density $\rhozero$ one obtains\,\cite{Piekarewicz:2008nh}:
\begin{subequations}
\begin{align}
 & {\cal E}_{\rm SNM}(\rho) = \epszero + \frac{1}{2}K_{0}x^{2}+\ldots ,\label{EandSa}\\
 & {\cal S}(\rho) = J + Lx + \frac{1}{2}K_{\rm sym}x^{2}+\ldots ,\label{EandSb}
\end{align} 
\label{EandS}
\end{subequations}
where $x\!=\!(\rho-\rhozero)\!/3\rhozero$ is a dimensionless parameter that quantifies the deviations of the 
density from its value at saturation. Here $\epszero$ and $K_{0}$ represent the energy per nucleon and the 
incompressibility coefficient of SNM. The linear term is absent because the pressure of symmetric nuclear
matter vanishes at saturation density. Thus, the small oscillations around the minimum energy $\epszero$
are controlled by the incompressibility coefficient $K_{0}$. The corresponding quantities in the case of the
symmetry energy are denoted by $J$ and $K_{\rm sym}$. However, unlike the case of symmetric nuclear 
matter, the slope of the symmetry energy $L$ does not vanish at saturation density. Indeed, assuming the 
validity of Eq.\,(\ref{SymmE}), $L$ is directly proportional to the pressure of pure neutron matter at saturation 
density:
\begin{equation}
   P_{0} \approx \frac{1}{3}\rhozero L \;.
 \label{PvsL}
\end{equation}
Hence, finding experimental observables that can effectively constrain the slope of the symmetry energy 
$L$ is tantamount to the determination of the pressure of a cold neutron gas at saturation density. As we 
show in Sec.\ref{sec:Results}, we will explore the predictions of several nuclear density functionals that 
while all successful in reproducing a host of laboratory observables, predict significant differences in the 
properties of neutron-rich systems, such as exotic nuclei and neutron stars.

\section{RESULTS}
\label{sec:Results}

This section is devoted to establish compelling connections between the properties of finite nuclei
and neutron stars. To assess uncertainties in the density dependence of the symmetry energy 
we rely on a set of nine successful covariant energy density functionals. Among them, 
NL3\cite{Lalazissis:1996rd,Lalazissis:1999} and IU-FSU\cite{Fattoyev:2010mx} have been used 
extensively in the literature. In particular, the IU-FSU functional represented an improvement over
the original FSUGold model\,\cite{Todd-Rutel:2005fa} by accounting for the existence of massive 
neutron stars\,\cite{Demorest:2010bx,Antoniadis:2013pzd,Cromartie:2019kug}. In addition, 
three different TAMU-FSU models, all with a relatively stiff symmetry energy, were introduced in 
Ref.\cite{Fattoyev:2013yaa} to explore whether existing experimental data could rule out thick 
neutron skins in ${}^{208}{\rm Pb}$. The remaining density functionals were calibrated for the 
first time using exclusively physical observables\,\cite{Chen:2014sca,Chen:2014mza}. That is, 
unlike earlier approaches, bulk properties of infinite nuclear matter were now predicted rather than 
assumed. Moreover, the calibration protocol relied on a statistically robust covariance analysis that 
provided both theoretical uncertainties and correlation coefficients\,\cite{Chen:2014sca}. The only 
significant difference in the calibration of these functionals was an assumed value for the presently 
unknown neutron skin thickness of $^{208}{\rm Pb}$\,\cite{Chen:2014mza}. 

\subsection{Ground State Properties}
\label{sec:GSProperties}

\begin{figure}[h]
\includegraphics[width=.95\linewidth]{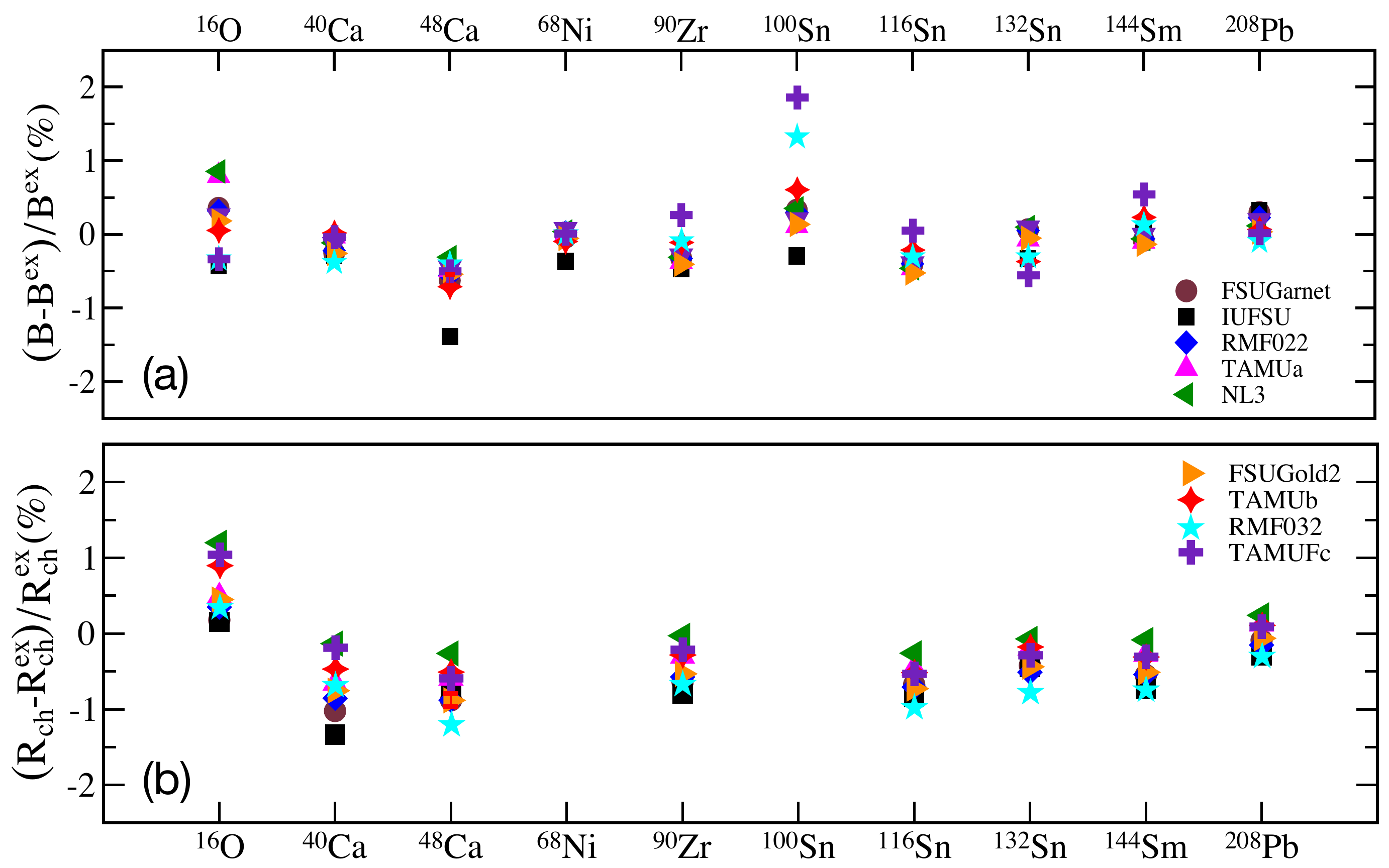}
\caption{Comparison between the theoretical predictions of the nine models introduced in the text 
for (a) the binding energy and (b) charge radius of a representative set of magic and semi-magic 
nuclei. Results are shown as the fractional discrepancy in percent.}
\label{Figure1}
\end{figure}

To assess the performance of the nine models employed in this work we display in Fig.\ref{Figure1} 
theoretical predictions relative to experiment for the binding energies per nucleon\,\cite{AME:2016} 
and charge radii\,\cite{Angeli:2013} of a representative set of magic and semi-magic nuclei. In all
cases the predictions fall within 2\% of the experimental values. However, it is worth mentioning
that for most of these functionals, the binding energies and charge radii displayed in the figure 
were incorporated into the fitting protocol. Nevertheless, these results suggest that extrapolations
to the high density regime characteristic of neutron stars involve covariant EDFs that are consistent 
with known properties of finite nuclei.

\subsection{Neutron Star Properties}
\label{sec:NSProperties}

Although both relativistic and non-relativistic energy density functionals have been enormously successful 
in describing ground-state properties of finite nuclei and their collective response, there is a distinct advantage 
in using a Lorentz covariant formulation as one extrapolates to dense nuclear matter. Inherent to any
consistent relativistic framework is the observance of ``causality", namely, the fact that no signal can propagate faster 
than the speed of light. In the context of dense matter, this implies a limit to the stiffness of the equation of state 
given by $P\!\le\!\Edens$, which in the context of Eq.(\ref{cs2}) implies that the speed of sound remains below 
the speed of light at all densities. However, the causal limit is often violated in non-relativistic descriptions, 
especially as central densities become large enough to support $2M_{\odot}$ neutron stars. Violating causality 
is particularly problematic in the case of the tidal polarizability as the relevant differential equation depends 
explicitly on the speed of sound; see Eq.(\ref{FandQ}).

\begin{figure}[ht]
\centering
 \includegraphics[width=.95\linewidth]{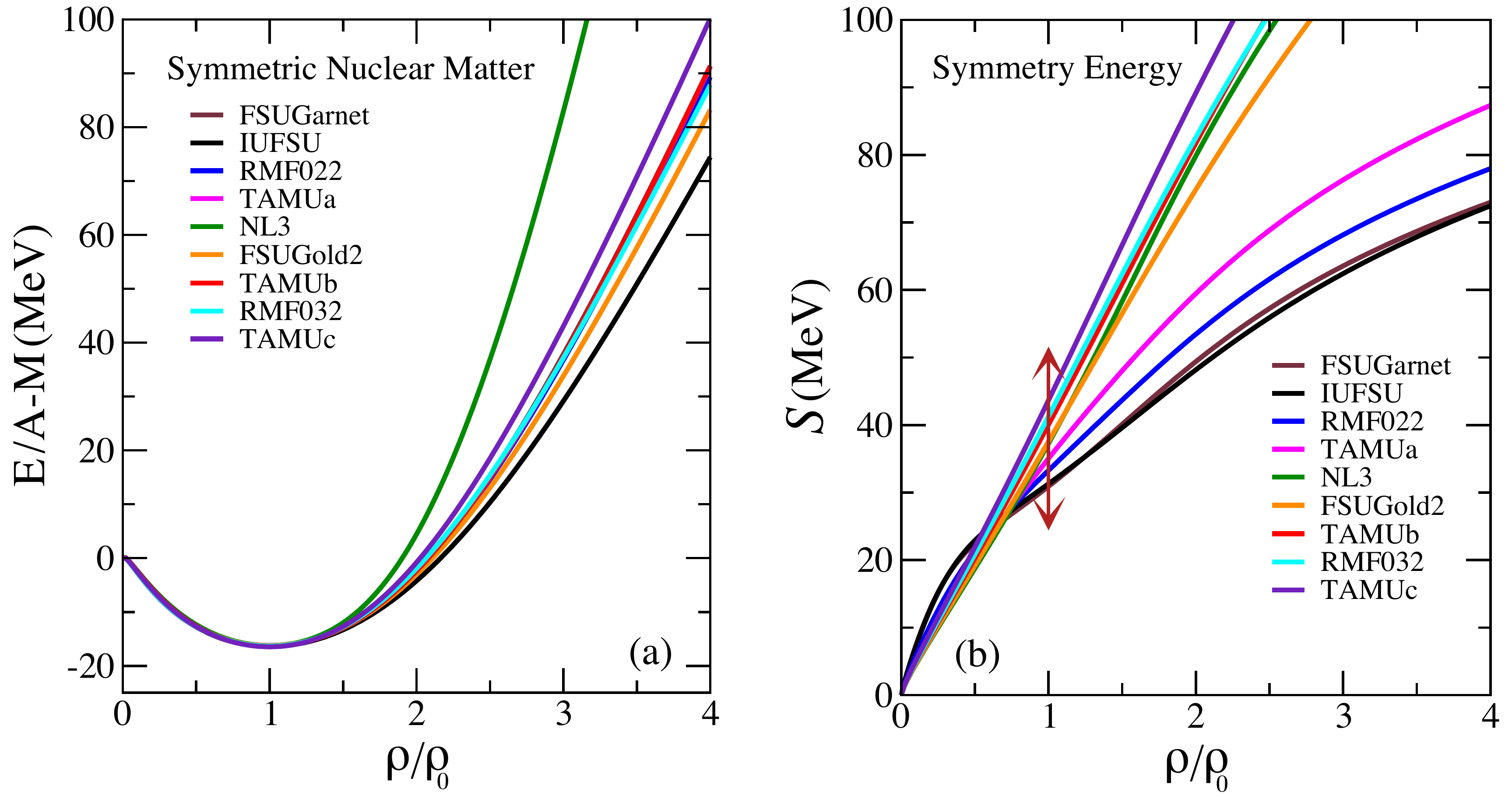}
\caption{Binding energy per nucleon (a) and nuclear symmetry energy (b) as a function of the baryon density
as predicted by the nine models described in the text. The arrow in (b) is indicative of the large model spread in the
slope of the symmetry energy at saturation density.}
\label{Figure2}
\end{figure}

Predictions for the equation of state of symmetric nuclear matter and the symmetry energy are displayed
in Fig.\ref{Figure2}. Under the assumption that Eq.(\ref{SymmE}) is valid, the EOS of pure neutron matter 
(not shown) is approximately equal to the sum of these two contributions. In the case of symmetric 
nuclear matter, all models predict a saturation point located at $\rhozero\!\approx\!0.15\,{\rm fm}^{-3}$ and 
a binding energy per nucleon of $\epszero\!\approx\!-16$\,MeV. Note that we use ``predict" as many of these 
functionals were calibrated using exclusively physical observables, namely, no bulk properties of nuclear-matter 
were incorporated into the calibration procedure\,\cite{Chen:2014sca,Chen:2014mza}. This suggest that the
values commonly adopted for both $\rhozero$ and $\epszero$ are properly encoded in certain bulk properties 
of finite nuclei.

Beyond the saturation point, the small oscillations around the minimum are controlled by the incompressibility 
coefficient $K_{0}$. Experimental measurements of the giant monopole resonance in ${}^{208}$Pb---and 
also on a few lighter nuclei such as ${}^{144}$Sm and ${}^{90}$Zr---have constrained the incompressibility coefficient 
to the $K_{0}\!=\!240\pm20$\,MeV range; see Ref.\cite{Garg:2018uam} and references contained therein. 
The NL3 model (shown in green) was conceived before such stringent constraints were available, leading
to a large incompressibility coefficient $K_{0}$ that, in turn, generates a very stiff EOS for symmetric nuclear
matter. In contrast, some of the most recently-calibrated functionals have incorporated for the first time 
information on giant monopole energies. As such, the incompressibility coefficient predicted by these 
models is fully consistent with experiment\,\cite{Chen:2014sca}. However, note that measurements of the 
distribution of isoscalar monopole strength in the isotopic chains of both tin and cadmium seem to suggest 
a smaller value for $K_{0}$\,\cite{Li:2007bp,Patel:2012zd}.  After more than a decade, the issue of the 
softness (or ``fluffiness") of these open-shell nuclei remains unresolved\,\cite{Garg:2006vc,Piekarewicz:2007us}. 

Whereas ground-state properties and collective excitations of finite nuclei impose stringent constraints on
the behavior of symmetric nuclear matter, this is no longer true for the symmetry energy; see 
Fig.\ref{Figure2}(b). It appears that nuclear ground-state properties---particularly the masses of 
neutron-rich nuclei---determine rather accurately the value of the symmetry energy at about two
thirds of nuclear matter saturation density, or at $\rho\!\approx\!(2/3)\rhozero\!\approx\!0.1\,{\rm fm}^{-3}$\,\cite{Horowitz:2000xj,Brown:2000,Furnstahl:2001un,Ducoin:2011fy,Horowitz:2014bja}. However, the slope 
of the symmetry energy in the vicinity of saturation density is poorly constrained by nuclear observables. 
In order to mitigate this problem, the neutron skin thickness of ${}^{208}$Pb was identified as an ideal proxy 
for $L$. Indeed, a very strong correlation was found between $L$ and the neutron skin thickness of 
${}^{208}$Pb\,\cite{Brown:2000,Furnstahl:2001un,Centelles:2008vu,RocaMaza:2011pm}. Given that 
symmetric nuclear matter saturates, the slope of the symmetry energy $L$ is directly related to the 
pressure of pure neutron matter at saturation density; see Eq.(\ref{PvsL}). As a result, a measurement 
of the neutron skin thickness of ${}^{208}$Pb provides critical information on a fundamental parameter 
of the equation of state. Motivated by this finding, the lead radius experiment (PREX) at JLab was 
commissioned about a decade ago and has already provided the first model-independent evidence in
favor of a neutron-rich skin in ${}^{208}$Pb\,\cite{Abrahamyan:2012gp,Horowitz:2012tj}. Unfortunately, 
due to unanticipated experimental challenges, PREX was not able to reach its original goal of a 1\% determination 
of the neutron radius of ${}^{208}$Pb. Since then, the follow-up PREX-II campaign was successfully completed 
and the brand new Calcium Radius EXperiment (CREX) was commissioned at the time of this 
writing\,\cite{CREX:2013}. In conjunction, PREX-II and CREX will provide valuable information on the 
equation of state of neutron-rich matter. Until then, one must explore how the uncertainties in the density 
dependence of the symmetry energy impact our predictions on the properties of neutron stars.

\begin{figure}[ht]
\centering
 \includegraphics[width=.55\linewidth]{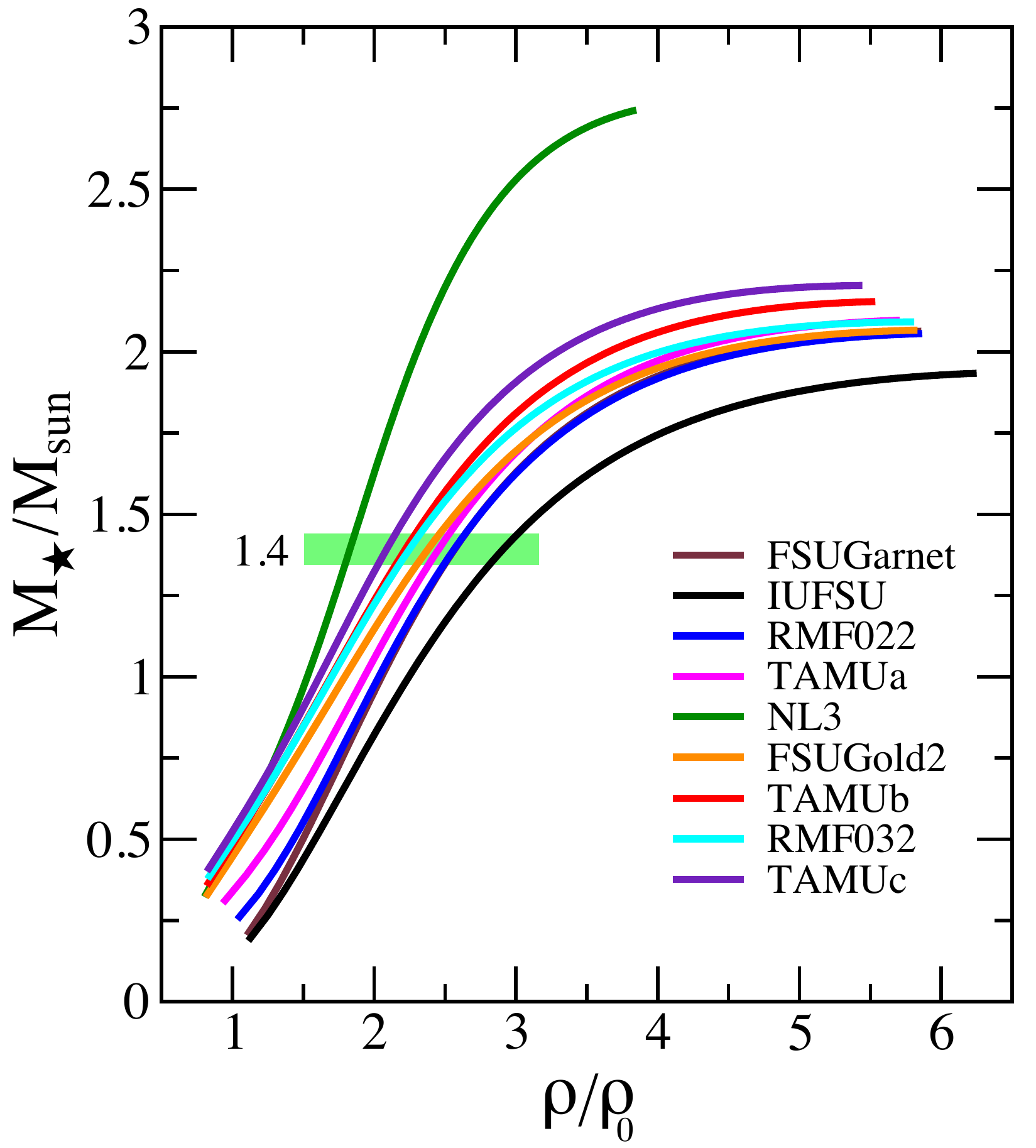}
\caption{Relationship between the mass of a neutron star and the central density required to support 
such a star as predicted by the nine models described in the text. The green bar illustrates the 
significant model dependence in the central density required to support a $1.4M_{\odot}$ neutron 
star.}
\label{Figure3}
\end{figure}

Although PREX-II and CREX constrain the behavior of neutron-rich matter in the vicinity of nuclear matter 
saturation density, neutron stars are sensitive to the equation of state up to several times saturation density. 
To assess the range of densities probed in the interior of neutron stars we display in Fig.\ref{Figure3} the 
central density required to support a neutron star of a given mass. As expected, the required central density 
depends critically on the stiffness of the equation of state. For example, in the case of NL3---the model with 
the stiffest EOS---the central density lies below  $4\rhozero$ for all masses below its predicted maximum 
mass of $\sim\!2.7M_{\odot}$. In contrast, the IUFSU model with the softest EOS requires a central density 
in excess of $6\rhozero$ to support a maximum mass of $\lesssim 2M_{\odot}$. Note that these densities 
may get even higher in the event of a phase transition in the stellar core---a situation that we do not contemplate
in this contribution. Finally, the green bar in the figure illustrates the model dependence in the central density 
that is required to support a ``canonical" $1.4 M_{\odot}$ neutron star: from less than twice $\rhozero$ (for NL3) 
to about three times $\rhozero$ for IUFSU. 

\begin{figure}[ht]
\centering
 \includegraphics[width=.99\linewidth]{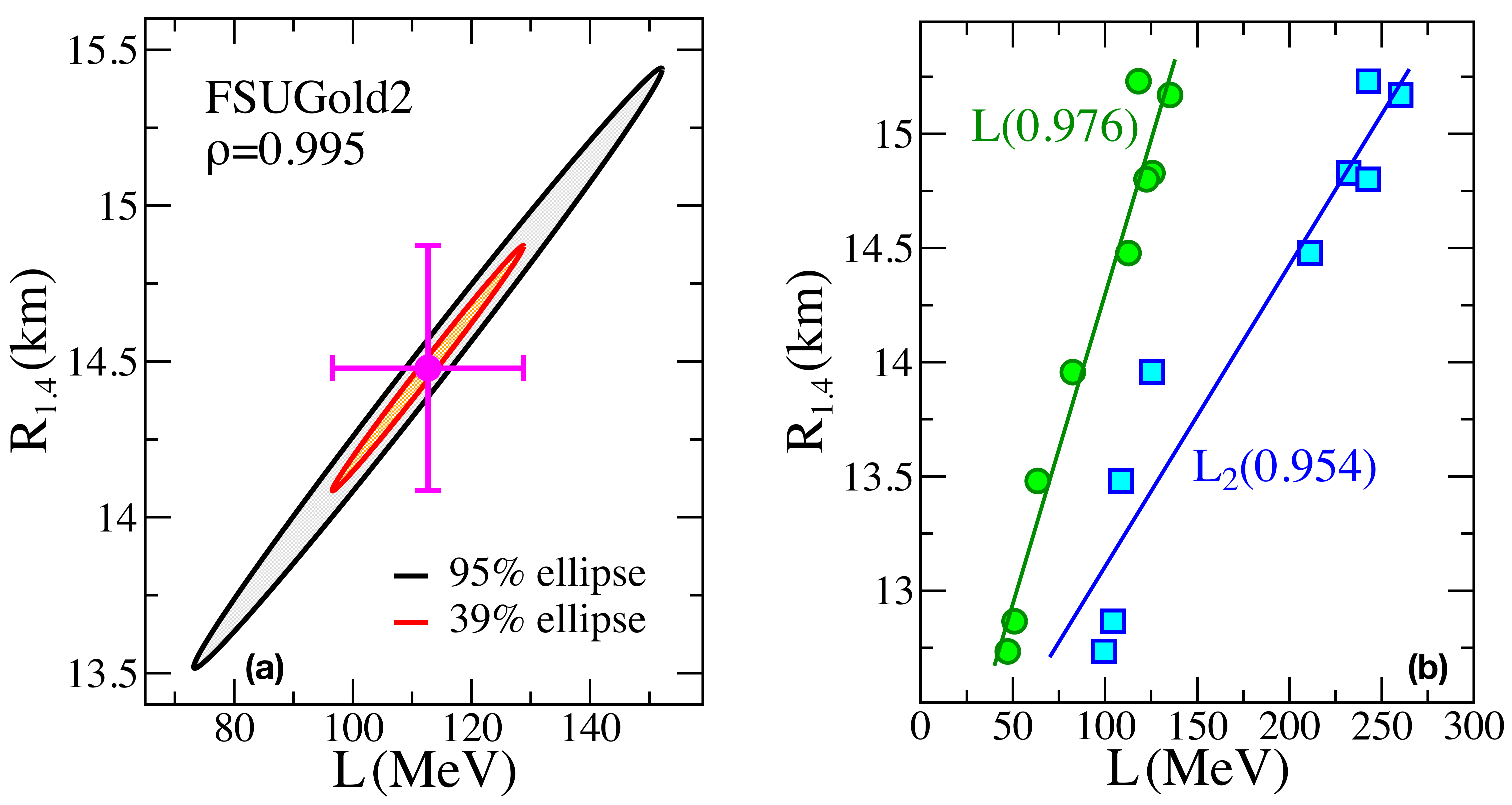}
\caption{The 39\% and 95\% confidence ellipses between the slope of symmetry energy $L$ and the radius 
of a $1.4M_{\odot}$ neutron star as predicted by the FSUGold2 density functional (a). Also displayed in the
figure are the corresponding statistical errors in $L$ and $R_{1.4}$. Systematic uncertainties in the same
correlation but now as predicted by the nine models described in the text (b). Also shown is the correlation 
between $L_2$ (the slope of symmetry energy at twice saturation density) and $R_{1.4}$.}
\label{Figure4}
\end{figure}

Stellar radii, however, seem to be largely determined by the density dependence of the symmetry energy 
in the immediate vicinity of nuclear matter saturation density. Indeed, it has been argued that the pressure 
near twice saturation density sets the overall scale for stellar radii\,\cite{Lattimer:2006xb}. This suggests 
that although PREX-II can not determine the stiffness of the EOS at high densities, it should provide valuable 
insights into the overall size of neutron stars\,\cite{Horowitz:2000xj,Horowitz:2001ya}. To underscores the 
strong correlation between the slope of the symmetry energy $L$ and the radius of a 
$1.4M_{\odot}$ neutron star we display in Fig.\,\ref{Figure4}(a) 39\% and 95\% confidence ellipses using
the FSUGold2 density functional as an example. FSUGold2 is particularly convenient to illustrate this 
correlation as no biases were introduced in the calibration of the functional---particularly in connection
to the (presently unknown) neutron skin thickness of ${}^{208}$Pb\,\cite{Chen:2014sca}. With a correlation
coefficient of almost one ($\rho\!=\!0.995$) and nearly ``degenerate" ellipses, a nearly one-to-one
correspondence exists between $L$ and $R_{1.4}$. Given that the neutron skin thickness of ${}^{208}$Pb
provides an ideal proxy for $L$, a powerful ``data-to-data" relation emerges between neutron-rich 
systems---finite nuclei and neutron stars---that differ in size by 18 orders of magnitude. 

Although the correlation displayed in Fig.\,\ref{Figure4}(a) is compelling, the statistical analysis carried out
is unable to assess \emph{systematic} errors associated to the intrinsic limitations of a given 
model; in this case FSUGold2. In order to properly assess systematic uncertainties, we include in
Fig.\,\ref{Figure4}(b) the predictions of each of the nine models considered in the text. Although slightly
weaker ($\rho\!=\!0.976$) than in Fig.\,\ref{Figure4}(a), the correlation between $L$ and $R_{1.4}$ remains
very strong. Note, however, that the correlation between $R_{1.4}$ and the slope of the symmetry energy 
at \emph{twice} saturation density ($L_{2}$) appears slightly weaker. In light of the expectation that stellar
radii are sensitive to the density dependence of the symmetry energy near twice saturation 
density\,\cite{Lattimer:2006xb}, our finding is mildly surprising, so it should be examined within the context
of a more diverse set of energy density functionals.

\begin{figure}[ht]
\centering
 \includegraphics[width=.99\linewidth]{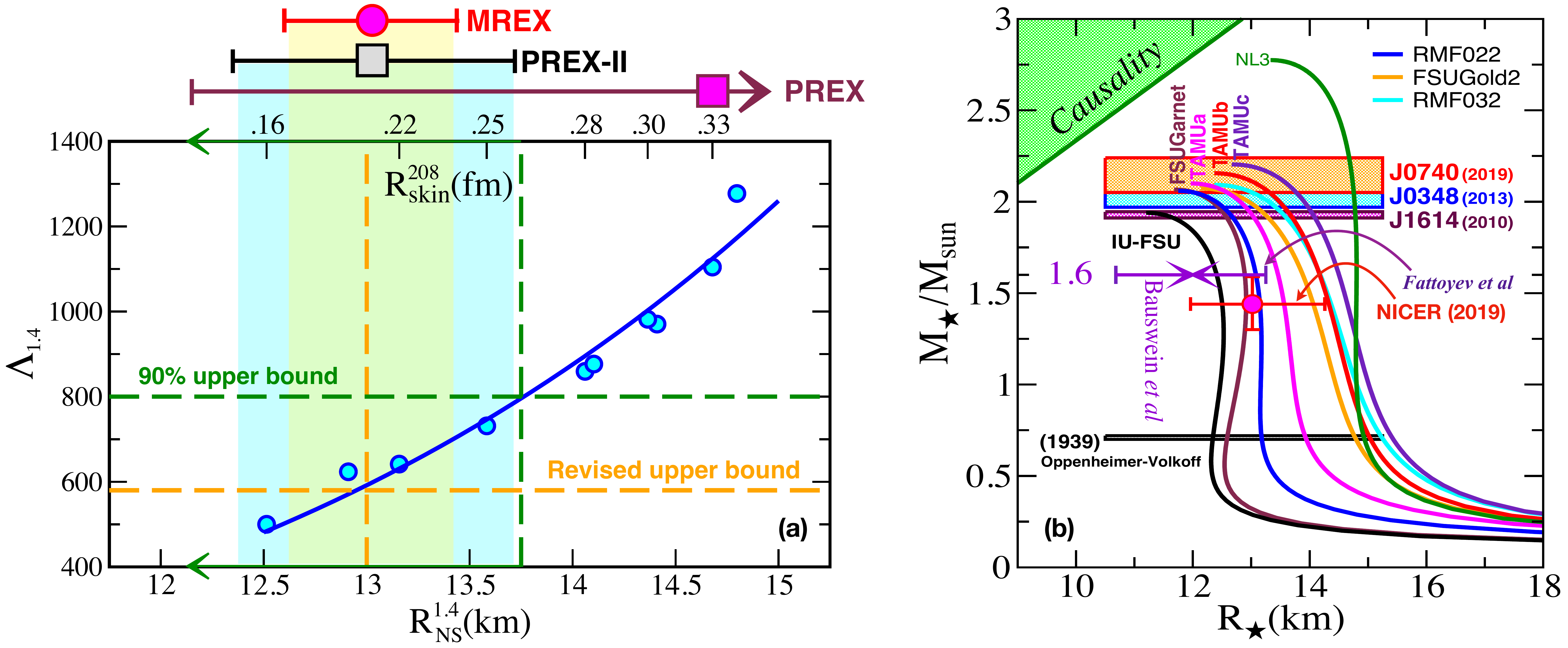}
\caption{(a) Tidal polarizability of a $1.4M_{\odot}$ neutron star as a function of both the stellar radius and 
the neutron skin thickness of $^{208}{\rm Pb}$ (a). The mass-vs-radius relation as predicted by the nine models 
described in the text, together with limits extracted from theory, electromagnetic observations, and gravitational 
wave detections (b). See text for further explanation.} 
\label{Figure5}
\end{figure}

We close this section by addressing the recent excitement in the field prompted by the historic detection of 
gravitational waves from the binary neutron star merger GW170817\,\cite{Abbott:PRL2017}. Unlike earlier
detections of black hole mergers that emit no electromagnetic radiation\,\cite{Abbott:PRL2016}, GW170817
opened the brand new era of multi-messenger astronomy. Indeed, the gravitational wave signal triggered 
public alerts that enabled myriad of telescopes operating at all wavelengths to follow the electromagnetic 
counterpart\,\cite{Drout:2017ijr,Cowperthwaite:2017dyu,Chornock:2017sdf,Nicholl:2017ahq}, a critical fact 
in establishing binary neutron star mergers as a favorable site for the formation of the heavy 
elements. Displayed in Fig.\ref{Figure5}(a) are predictions for the dimensionless tidal polarizability 
$\Lambda$ of a $1.4 M_{\odot}$ neutron star as a function of the stellar radius\,\cite{Fattoyev:2017jql,
Thiel:2019tkm,Piekarewicz:2019ahf}. Given the strong sensitivity of $\Lambda$ to the stellar compactness 
as indicated in Eq.(\ref{Lambda}), the displayed correlation is very strong once the stellar mass has been fixed. Indeed,
since $k_{2}$ is known to display a mild sensitivity to the underlying equation of state\,\cite{Piekarewicz:2018sgy}, 
the curve fitted to the theoretical predictions scales approximately with the fifth power of the 
radius\,\cite{Fattoyev:2017jql}. Shown in the upper abscissa is the PREX result with its associated large
error bar\,\cite{Abrahamyan:2012gp,Horowitz:2012tj}, alongside the anticipated more precise determinations
from PREX-II and MREX\,\cite{Thiel:2019tkm}. Note that while the error bars are realistic, the central values 
are placed arbitrarily at $R_{\rm skin}^{208}\!\simeq\!{0.2}\,{\rm fm}$. 

The extraction of the tidal polarizability of a $1.4 M_{\odot}$ neutron star provides the strongest constraint
from GW170817 on the EOS of neutron-rich matter. In the initial discovery paper\,\cite{Abbott:PRL2017}, the 
LIGO-Virgo collaboration placed a 90\% upper bound of $\Lambda_{1.4}\!\le\!800$ that was stringent enough 
to disfavor overly stiff EOSs\,\cite{Fattoyev:2017jql,Annala:2017llu}. Since then, some of the initial assumptions 
have been relaxed leading to the more stringent upper limit of $\Lambda_{1.4}\!=\!190^{+390}_{-120}$, implying 
a common radius for the two colliding neutron stars of $R\!=\!11.9\pm1.4\,{\rm km}$\,\cite{Abbott:2018exr}. As 
illustrated in the figure, this revised upper bound creates enormous tension as it excludes most theoretical 
models---even when \emph{all} the models provide an excellent description of the ground state properties 
of finite nuclei. 

The tension is further exacerbated as one examines masses and radii simultaneously. The characteristic
mass-radius relation as predicted by the nine models used in the text is displayed on Fig.\ref{Figure5}(b). 
In addition to these nine theoretical predictions the figure includes several interesting limits. The 1939 
prediction by Oppenheimer and Volkoff for the maximum neutron star mass---assuming that the entire 
pressure support is due to a non-interacting Fermi gas of neutrons---is displayed in the lower part of the 
figure\,\cite{Opp39_PR55}. This pioneering prediction has long been refuted, especially with the confirmation 
of three neutron stars with masses in the vicinity of 
2$M_{\odot}$\,\cite{Demorest:2010bx,Antoniadis:2013pzd,Cromartie:2019kug}; see the three bars in
the upper portion of the figure. In particular, Cromartie and collaborators have measured a neutron star with 
a mass of about 2.14$M_{\odot}$\,\cite{Cromartie:2019kug}---a value that is tantalizingly close to the 
\emph{upper limit} of $M_{\rm max}\!=\!2.17M_{\odot}$ suggested by Margalit and Metzger from exploiting 
the multimessenger nature of GW170817\,\cite{Margalit:2017dij}. By also combining gravitational-wave and
electromagnetic information from GW170817, Bauswein and collaborators provided a \emph{lower limit} on 
the radius of a $1.6 M_{\odot}$ neutron star\,\cite{Bauswein:2017vtn} which, when combined with the upper
limits obtained in Refs.\,\cite{Fattoyev:2017jql,Annala:2017llu}, results in the two arrows facing each other in 
the figure. Finally, the figure includes results from the very recent (few days old!) simultaneous extraction of 
the mass and radius of PSR J0030+0451 by the Neutron Star Interior Composition Explorer (NICER). The
quoted results by Miller and collaborators are:  $M\!=\!1.44^{+0.15}_{-0.14}M_{\odot}$ and
$R\!=\!13.02^{+1.24}_{-1.06}M_{\odot}$\,\cite{Miller:2019cac}. These values are 
consistent with the independent analysis reported by Riley and collaborators in Ref.\,\cite{Riley:2019yda}.
Although the first NICER results do not impose stringent constraints on the EOS, this pioneering 
measurement determined for the first time the gravitational mass and equatorial radius of a neutron star.

So what do we conclude? On the one hand, the existence of massive neutron stars suggests that the EOS 
at high densities must be relatively stiff to provide the necessary pressure support. On the other hand, 
GW170817 seems to favor compact stars with small radii---suggesting instead that the EOS must be
soft. How can we then simultaneously account for both small radii and large masses? As argued earlier, 
stellar radii appear to be sensitive to the EOS of neutron-rich matter in the vicinity of nuclear matter 
saturation density. In contrast, the maximum neutron star mass is sensitive to the equation of state at the 
highest densities attained in the stellar core. Hence, the apparent tension may be resolved if the EOS is 
soft at intermediate densities---thereby accounting for the small radii---but then stiffens at higher densities 
in order to support heavy neutron stars. This already unique situation could become even more interesting if 
PREX-II confirms the original PREX measurement of a neutron skin thickness of 
$R_{\rm skin}^{208}\!=\!{0.33}\,{\rm fm}$, albeit with larger error bars\,\cite{Abrahamyan:2012gp,Horowitz:2012tj}. 
If confirmed, this would imply that the EOS is stiff in the vicinity of saturation density, it will then soften at 
intermediate densities to account for the small stellar radii, but will ultimately stiffen at high densities to 
explain the existence of massive neutron stars. The evolution from stiff to soft and back to stiff may reflect 
a fascinating underlying dynamics, perhaps indicative of an exotic phase transition in the stellar interior.

\section{CONCLUSIONS}
\label{sec:Conclusions}

Nuclear science is driven by the quest to understand the fundamental interactions that shape the structure 
of the universe. A new generation of terrestrial facilities being commissioned all over the world will help 
answer some key science questions, such as \emph{How did visible matter come into being and how 
does it evolve?} and \emph{How does subatomic matter organize itself and what phenomena 
emerge?}\,\cite{LongRangePlan}. Insights into the dynamics of neutron-rich matter will emerge as one 
probes exotic nuclei with very large neutron skins. In the cosmos, neutron-rich matter is at the heart of many 
fundamental questions that include: \emph{What are the new states of matter at exceedingly high density 
and temperature?} and \emph{How were the elements from iron to uranium made?}\,\cite{NAP10079}. 
Remarkable development within the last few years---and in some cases during the past few months---are 
providing valuable insights into the nature of dense neutron-rich matter. First, the direct detection of
gravitational waves from the binary neutron star merger GW170817 suggests that neutron stars are 
fairly compact, implying a relatively soft EOS at intermediate densities\,\cite{Abbott:PRL2017}. Second, 
the observation by Cromartie and collaborators of the most massive neutron star to date implies that
the EOS must stiffen at high densities\,\cite{Cromartie:2019kug}. Finally, NICER---aboard the international
space station---reported the very first simultaneous measurement of the mass and radius of a neutron 
star\,\cite{Miller:2019cac,Riley:2019yda}. This pioneering result is highly significant as a one-to-one
correspondence exists between the mass-radius relation of neutron stars and the underlying equation 
of state\,\cite{Lindblom:1992}.

As we embark on this new journey of discovery, nuclear theory will play a critical role in guiding new 
experimental programs. As critical, nuclear theory will continue to make predictions in regimes that 
will remain inaccessible to experiment and observation. Prospects in nuclear theory are excellent 
given the recent advances in ab initio methods that start from chiral EFT Hamiltonians fitted to 
two- and three-body data\,\cite{Furnstahl:2019lue}. Indeed, within the last decade ab initio calculations 
have seen an explosive growth in scalability to larger systems. Yet despite this undeniable progress,
density functional theory remains the most promising and only tractable approach that may be applied 
over the entire nuclear landscape: from finite nuclei to neutron stars. It was the main goal of this review
to demonstrate the power and flexibility of modern covariant energy density functionals in predicting the
properties of nuclear system across such a rich and diverse landscape. Particularly important in this
context is the unique synergy between nuclear physics and astrophysics in the brand new era of 
gravitational wave astronomy.

So what is the path forward in the development of density functional theory as it pertains to nuclear 
physics? Perhaps the most serious obstacle is the lack of a one-to-one correspondence between the 
one-body nuclear density and a suitable external potential, a requirement that is germane to DFT as 
originally conceived by Hohenberg and Kohn\,\cite{Hohenberg:1964zz,Kohn:1965}. Moreover, unlike
DFT applications to electronic structure where the fundamental interaction is known, the underlying 
nucleon-nucleon interaction---although often inspired by QCD---relies on fits to two- and three-nucleon 
data. A much more fruitful application of DFT to nuclear physics is through the Kohn-Sham 
equations, a set of equations that are highly reminiscent of the traditional mean-field approach 
that lies at the heart of nuclear physics. However, in contrast to the Kohn-Sham 
formalism that yields in principle the exact ground-state energy and one-body density, no such
guarantee exists in nuclear physics since the ``universal" nuclear mean-field potential is unknown.
Nevertheless, enormous progress in ab initio approaches provide meaningful benchmarks for the 
refinement of existing nuclear functionals. The CREX campaign at JLab was motivated in part by
the powerful connection between ab initio approaches and DFT\,\cite{CREX:2013,Horowitz:2013wha}.
Finally, nuclear density functionals will be informed and refined by the wealth of experimental and 
observational data that will emerge from rare isotope facilities, telescopes operating across the
entire electromagnetic spectrum, and ever more sensitive gravitational wave detectors. This
unique synergy will prove vital in our quest to determine the nuclear equation of state.

\section*{ACKNOWLEDGMENTS}
This material is based upon work supported by the U.S. Department of Energy Office of Science, 
Office of Nuclear Physics under Award Number DE-FG02-92ER40750. 

\bibliographystyle{ar-style5}
\bibliography{RelativisticDFT.bbl}

\vfill\eject
\end{document}